\documentclass[12pt,aps,prb,preprint]{revtex4}
\usepackage{amsfonts}
\usepackage{amsmath}
\usepackage{amssymb}
\usepackage{graphicx}

\input{tcilatex}

\begin{document}

\title{Unfamiliar trajectories for a relativistic particle in a Kepler or
Coulomb potential $V(r)=-\alpha /r$}
\author{Timothy H. Boyer}
\affiliation{Department of Physics, City College of the City University of New York, New
York, New York 10031}
\pacs{PACS number}

\begin{abstract}
Relativistic particles in the Kepler and Coulomb potentials may have
trajectories that are qualitatively different from the trajectories found in
nonrelativistic mechanics. Spiral scattering trajectories were pointed out
by C. G. Darwin in 1913 in connection with the relativistic Rutherford
scattering of classical charged particles. Relativistic trajectories are of
current interest in connection with Cole and Zou's computer simulation of
the hydrogen ground state in classical physics.
\end{abstract}

\maketitle

%\keywords{relativistic trajectories}

\section{Introduction}

The mechanics of a relativistic particle in a potential is usually treated
as an afterthought in classical mechanics textbooks. These textbooks usually
begin with an extensive treatment of nonrelativistic mechanics, then the
energy and momentum of relativistic particles are noted, and the corrections
to the orbits of nonrelativistic particles are sometimes mentioned.\cite%
{Goldstein} However, relativistic particles in the familiar Kepler or
Coulomb potential $V(r)=-\alpha /r$ can have trajectories that are
qualitatively different from the trajectories found from nonrelativistic
mechanics, and these unfamiliar trajectories are not mentioned in the
textbooks.\cite{Darwin} For example, a relativistic particle in a $1/r$
potential can spiral into the force center (while conserving mechanical
energy and angular momentum). This behavior occurs because a small increase
in the velocity near the speed of light $c$ can lead to a large increase in
the mass $m/(1-v^{2}/c^{2})^{1/2}$ so that an increase in the kinetic energy
will compensate the decrease in the potential energy as the radius
decreases, thus keeping the total energy constant. In addition, the increase
in the linear momentum will compensate the decrease in the radius to keep
the angular momentum constant. The existence of such trajectories has played
a role in recent research.\cite{Cole,Boyer}

The potential $V(r)=-\alpha /r$ ($\alpha >0$) appears in the Kepler problem
of gravitational physics and in the attraction of point charges in
electrostatics. The theory of gravitation finds its relativistic form in
general relativity and electrostatics has a natural extension into
electrodynamics. However, we will not discuss these extended theories. The
problem we consider has no curved spacetime and no radiation; it is the
relativistic mechanics of a particle in a $1/r$ potential.

As a first example of the unfamiliar nature of some relativistic particle
trajectories, we show that if the angular momentum is too small, there are
no circular orbits for a relativistic particle in a $1/r$ potential. In
contrast, in nonrelativistic mechanics, such a situation never occurs; there
are always nonrelativistic circular orbits unless the angular momentum
vanishes.

We follow the traditional procedure for the classification of trajectories
for a particle, noting the orbits that undergo a qualitative transformation
in the nonrelativistic limit and those that do not. The equations of the
trajectories are then obtained. We find the familiar scattering trajectories
and the familiar rosette shapes that appear in texts on old quantum theory,%
\cite{Ruark} and also C. G. Darwin's spiraling trajectories,\cite{Darwin}
which do not appear in textbooks. Finally, we mention the research context
in which these spiral trajectories have recently become of interest.

\section{The Mechanical Problem}

We compare the nonrelativistic and relativistic behavior of a particle in
the potential $V(r)=-\alpha /r$. A particle in such a potential experiences
a force $\mathbf{F=-\nabla }V(r\mathbf{)}=-\hat{r}\alpha /r^{2}$. If we use
the nonrelativistic momentum $m\mathbf{v}$ for a nonrelativistic particle,
Newton's second law becomes 
\begin{equation}
\frac{d}{dt}(m\mathbf{v})=-\frac{\alpha }{r^{2}}\hat{r}.  \label{1}
\end{equation}%
For a relativistic particle, the momentum is $m\mathbf{v}/\sqrt{1-v^{2}/c^{2}%
}$, which gives 
\begin{equation}
\frac{d}{dt}\big(\frac{m\mathbf{v}}{\sqrt{1-v^{2}/c^{2}}}\big)=-\frac{\alpha 
}{r^{2}}\hat{r}.  \label{2}
\end{equation}%
If we take the dot product of Eqs.~(\ref{1}) and (\ref{2}) with the velocity 
$\mathbf{v}$ and integrate with respect to time, we obtain conservation of
energy. \ \ The nonrelativistic energy $\mathcal{E}_{nr}$\ is expressed as 
\begin{equation}
\mathcal{E}_{nr}=\frac{1}{2}mv^{2}-\frac{\alpha }{r},
\end{equation}%
while the relativistic energy $E$ is 
\begin{equation}
E=\mathcal{E}+mc^{2}=\frac{mc^{2}}{\sqrt{1-v^{2}/c^{2}}}-\frac{\alpha }{r}.
\label{4}
\end{equation}%
Here we have written $E=\mathcal{E}+mc^{2}$ so that $\mathcal{E}$ is the
energy difference from the particle rest energy. Also, because the $1/r$
potential gives a central force, the angular momentum $L$ is conserved,
giving in the nonrelativistic case 
\begin{equation}
\mathbf{L}_{nr}\mathbf{=r\times (}m\mathbf{v)},
\end{equation}%
and in the relativistic case 
\begin{equation}
\mathbf{L=r\times }\big(\frac{m\mathbf{v}}{\sqrt{1-v^{2}/c^{2}}}\big).
\end{equation}%
We will use the conservation laws to determine the trajectories. As an
introduction to the nature of the differences that appear in relativistic
mechanics, we first consider the case of circular orbits.

\section{Limiting Angular Momentum for Relativistic Circular Orbits}

For circular orbits, the particle displacement $\mathbf{r}$ from the center
of the potential is perpendicular to the velocity $\mathbf{v}$, and the
angular momentum $\mathbf{L}$ is perpendicular to the plane of the orbit.
The magnitude of the angular momentum in the nonrelativistic case is 
\begin{equation}
L_{nr}=mrv,
\end{equation}%
and in the relativistic case 
\begin{equation}
L=\frac{mrv}{\sqrt{1-v^{2}/c^{2}}}.
\end{equation}%
For circular orbits in the nonrelativistic case, we have 
\begin{equation}
m\frac{v^{2}}{r}=\frac{\alpha }{r^{2}},
\end{equation}%
and in the relativistic case 
\begin{equation}
\frac{m}{\sqrt{1-v^{2}/c^{2}}}\frac{v^{2}}{r}=\frac{\alpha }{r^{2}}.
\end{equation}%
If we now use the angular momentum to remove either $r$ or $v$, we find for
the nonrelativistic case 
\begin{equation}
v=\frac{\alpha }{L_{nr}}\quad \text{or}\quad r=\frac{L_{nr}^{2}}{m\alpha }.
\end{equation}%
Because both $v$ and $r$ can take all values between zero and infinity in
nonrelativistic mechanics, it follows that any value of the angular momentum 
$L_{nr}$ can lead to a circular orbit.

The same procedure in the relativistic case leads to 
\begin{subequations}
\label{eqrelv}
\begin{eqnarray}
v &=&\frac{\alpha }{L},  \label{12a} \\
\noalign{\noindent or}r &=&\frac{L^{2}}{m\alpha }\big[1-\big(\frac{\alpha }{%
Lc}\big)^{2}\big]^{1/2}.  \label{12b}
\end{eqnarray}%
In the relativistic case, we have an upper limit for the speed $v=c$, and
accordingly a lower limit on the magnitude of the angular momentum for a
circular orbit. As seen in Eq.~(\ref{eqrelv}), we must have\cite{Torkellson} 
\end{subequations}
\begin{equation}
L>\frac{\alpha }{c}  \label{13}
\end{equation}%
Thus there is a qualitative distinction for circular orbits between
nonrelativistic and relativistic mechanics for the potential $V(r)=-\alpha
/r $, and only for this potential.\cite{only} The limiting value of the
angular momentum in the relativistic case is $L_{\alpha }=\alpha /c$. In
contrast, in nonrelativistic mechanics there is no limiting speed and thus
no lower limit on the angular momentum for circular orbits.

The actual value of the limiting angular momentum $L_{\alpha }$ for the $1/r$
potential depends on the magnitude of $\alpha $. For an electron of charge $%
e $ in the field of a nucleus of atomic number $Z$, $\alpha =Ze^{2}$, and
the limiting angular momentum can be written as $L_{\alpha
}=Zm(e^{2}/mc^{2})c=Z(e^{2}/\hbar c)\hbar $. Thus for an electron in the
Coulomb potential of hydrogen , the limiting angular momentum $L_{\alpha
}=(e^{2}/\hbar c)\hbar \simeq (1/137)\hbar $. However the reader should not
be distracted by an angular momentum that is close to that given by Planck's
constant. Our discussion involves only relativistic classical mechanics, not
quantum mechanics.

\section{Classification of Trajectories}

The classification of orbits for the potential $V(r)=-\alpha /r$ reflects
the fact that some relativistic trajectories are qualitatively different
from those found in nonrelativistic mechanics. The classification can be
made following the usual procedures of classical mechanics.\cite{Goldstein}
The motion is confined to a plane so that the velocity can be written in
terms of polar coordinates as $\mathbf{r}=\hat{r}\dot{r}+\hat{\theta}r\dot{%
\theta}$. Then we use the energy expression (4) and remove the $\dot{\theta}$
dependence in the velocity in favor of the angular momentum $L$ to obtain a
first-order differential equation in the radial variable $r$ as a function
of time. We will carry out the procedure for the relativistic case and then
take the nonrelativistic limit to connect with the familiar Kepler orbits.

In terms of polar coordinates, the relativistic angular momentum is 
\begin{equation}
L=\frac{mr^{2}\dot{\theta}}{\sqrt{1-(\dot{r}^{2}+r^{2}\dot{\theta}^{2})/c^{2}%
}},  \label{eq:relL}
\end{equation}%
where we choose the orientation so that $L$ and $\dot{\theta}$ are positive.
If we solve Eq.~(\ref{eq:relL}) for $\dot{\theta}$ and substitute the result
into Eq.~(\ref{4}), we find 
\begin{equation}
E=\mathcal{E}+mc^{2}=\frac{mc^{2}}{\sqrt{1-(\dot{r}^{2}/c^{2})-L^{2}(1-\dot{r%
}^{2}/c^{2})/(m^{2}r^{2}c^{2}+L^{2})}}-\frac{\alpha }{r}.  \label{eq:ee}
\end{equation}%
The solution of Eq.~(\ref{eq:ee}) for $\dot{r}^{2}$ is 
\begin{equation}
\dot{r}^{2}=c^{2}\bigg[1-\big(1+\frac{L^{2}}{m^{2}r^{2}c^{2}}\big)\big(\frac{%
mc^{2}}{E+(\alpha /r)}\big)^{2}\bigg].  \label{17}
\end{equation}

In the following, it will turn out that the limiting angular momentum, $%
L_{\alpha }=\alpha /c$, that appears in the analysis of circular orbits,
represents a crucial limiting value. It is useful to have certain
constraints on the energy which are associated with this limiting angular
momentum. Because the particle velocity is less than $c$, we see from Eqs.~(%
\ref{eq:relL}) and (\ref{eq:ee}) that 
\begin{equation}
L=\frac{mr^{2}\dot{\theta}}{\sqrt{1-v^{2}/c^{2}}}<\frac{rmc}{\sqrt{%
1-v^{2}/c^{2}}}=\frac{r}{c}\big(E+\frac{\alpha }{r}\big).  \label{eq:this}
\end{equation}%
Equation~(\ref{eq:this}) requires that 
\begin{equation}
L<\frac{r}{c}\big(E+\frac{\alpha }{r}\big)\quad \text{or \ \ }L-\frac{\alpha 
}{c}<\frac{Er}{c}.  \label{19}
\end{equation}%
Because $r$ is positive, it follows that if $L\geq \alpha /c$, then $E=%
\mathcal{E}+mc^{2}>0$. Thus orbits with angular momentum larger than $%
L_{\alpha }$ must have positive total (relativistic) energy. In particular,
bound orbits with $L>L_{\alpha }$ (those that do not plunge into the
potential center) cannot have arbitrarily small values of kinetic plus
potential energy; rather the kinetic plus potential energy $\mathcal{E}$
must be larger than $-mc^{2}$.

The classification of the relativistic orbits can be made by noting that the
function on the right-hand side of Eq.~(\ref{17}) must satisfy the
requirement $0\leq \dot{r}^{2}<c^{2}$. It is straightforward to transform
this condition to 
\begin{equation}
-L^{2}c^{2}<0\leq (E^{2}-m^{2}c^{4})r^{2}+2E\alpha r+(\alpha
^{2}-L^{2}c^{2}).  \label{21}
\end{equation}%
Because all quantities are real, the first inequality in Eq.~(\ref{21})
holds automatically and is not important. The second inequality can be
treated by plotting the parabolic function $Y(r)=(E^{2}-m^{2}c^{4})r^{2}+2%
\alpha Er+(\alpha ^{2}-L^{2}c^{2})$ versus $r$. The allowed orbits
correspond to the regions of positive $r$ where $Y(r)>0$ and the turning
points occur when $Y(r)=0$: 
\begin{equation}
r_{\mathrm{turning-point}}=\frac{E\alpha \pm \sqrt{E^{2}\alpha
^{2}+(m^{2}c^{4}-E^{2})(\alpha ^{2}-L^{2}c^{2})}}{m^{2}c^{4}-E^{2}}.
\label{22}
\end{equation}%
In the nonrelativistic limit $c\rightarrow \infty $ with $E-mc^{2}=\mathcal{%
E\rightarrow E}_{nr}$, the inequality (\ref{21}) becomes 
\begin{equation}
-L_{nr}^{2}/m<0\leq \mathcal{E}_{nr}r^{2}+\alpha r-L_{nr}^{2}/2m,
\end{equation}%
where the first inequality holds for any real $L_{nr}$. The parabolic
function $y(r)=\mathcal{E}_{nr}r^{2}+\alpha r-L_{nr}^{2}/2m$ can be plotted
versus $r$ with the allowed orbits corresponding to regions of $r>0$ and $%
y(r)>0$; the turning points occur when $y(r)=0$, 
\begin{equation}
r_{\mathrm{turning-point}}=\frac{-\alpha }{2\mathcal{E}_{nr}}\Big(1\pm \sqrt{%
1+\frac{2L_{nr}^{2}\mathcal{E}_{nr}}{m\alpha ^{2}}}\Big).  \label{24}
\end{equation}

A relativistic circular orbit corresponds to the square root in Eq.~(\ref{22}%
) equal to zero so that both the inner and outer turning points are at the
same radius $r$. The vanishing square root gives a connection between $E$
and $L$ for a circular orbit, 
\begin{equation}
E=mc^{2}\sqrt{1-\Big(\frac{\alpha }{Lc}\Big)^{2}},  \label{25}
\end{equation}%
If we substitute Eq.~(\ref{25}) into Eq.~(\ref{22}), we obtain a radius $r$,
which is the same result obtained in Eq.~(\ref{12b}). Also, if we take the
nonrelativistic limit of Eq.~(\ref{25}), we obtain 
\begin{equation}
\mathcal{E}=E-mc^{2}=mc^{2}\sqrt{1-\Big(\frac{\alpha }{Lc}\Big)^{2}}%
-mc^{2}\cong -\frac{1}{2}\frac{m\alpha ^{2}}{L_{nr}^{2}}=\mathcal{E}_{nr},
\label{26}
\end{equation}%
which is the nonrelativistic expression for a circular orbit. We note that
Eq.~(\ref{25}) gives a limit $E\geq 0$ for a circular orbit, which agrees
with the result given below Eq.~(\ref{19}). However, there is no lower bound
for the energy in the nonrelativistic approximation appearing in Eq.~(\ref%
{26}) when we take the nonrelativistic limit $L_{nr}\rightarrow 0$ (and find 
$\mathcal{E}_{nr}\mathcal{\rightarrow }-\infty $).

The general character of the turning points in the nonrelativistic equation (%
\ref{24}) is controlled by the sign of the single parameter $\mathcal{E}%
_{nr} $, with $L_{nr}=0$ a unique special case. In the full relativistic
treatment, the character of the turning points in Eq.~(\ref{22}) depends on
the signs of both $E\alpha /(m^{2}c^{4}-E^{2})$ and $(m^{2}c^{4}-E^{2})(%
\alpha ^{2}-L^{2}c^{2})$, where there is a sign change in the second term
depending on the magnitude of $L$. When $L>L_{\alpha }$ so that $(\alpha
^{2}-L^{2}c^{2})<0$, the turning point analysis is qualitatively the same
for both the relativistic and nonrelativistic cases. However, if the angular
momentum is small but non-zero, $0<L\leq \alpha /c$, so that $(\alpha
^{2}-L^{2}c^{2})$ is positive, then the turning point analysis allows new
possibilities which do not appear in the nonrelativistic case. We will see
these new trajectories when we obtain the orbit equations.

\section{Orbit Equations}

The orbit equations $r(\theta )$ can be found using the traditional
analysis. If we write the relativistic particle momentum in the form 
\begin{equation}
\mathbf{p=}\hat{r}p_{r}+\hat{\theta}p_{\theta }=\frac{m(\hat{r}\dot{r}+\hat{%
\theta}r\dot{\theta})}{\sqrt{1-v^{2}/c^{2}}}=\frac{m\hat{r}\dot{r}}{\sqrt{%
1-v^{2}/c^{2}}}+\hat{\theta}\frac{L}{r},  \label{27}
\end{equation}%
then the relativistic particle energy $E$ can be written as 
\begin{equation}
E=\sqrt{p^{2}c^{2}+m^{2}c^{4}}-\frac{\alpha }{r},
\end{equation}%
or 
\begin{equation}
\big(E+\frac{\alpha }{r}\big)^{2}=p^{2}c^{2}+m^{2}c^{4}=p_{r}^{2}c^{2}+\frac{%
L^{2}c^{2}}{r^{2}}+m^{2}c^{4}.  \label{29}
\end{equation}%
Now from Eq.~(\ref{27}) 
\begin{equation}
\frac{p_{r}}{p_{\theta }}=\frac{\dot{r}}{r\dot{\theta}}=\frac{1}{r}\frac{dr}{%
d\theta },
\end{equation}%
and 
\begin{equation}
p_{r}=\frac{L}{r^{2}}\frac{dr}{d\theta }.
\end{equation}%
Thus Eq.~(\ref{29}) becomes 
\begin{equation}
\big(E+\frac{\alpha }{r}\big)^{2}=\big(\frac{L}{r^{2}}\frac{dr}{d\theta }%
\big)^{2}c^{2}+\frac{L^{2}c^{2}}{r^{2}}+m^{2}c^{4}.  \label{32}
\end{equation}%
If we introduce the usual $r$-inverse variable, $s=1/r$, so that 
\begin{equation}
\frac{ds}{d\theta }=-\frac{1}{r^{2}}\frac{dr}{d\theta },
\end{equation}%
Eq.~(\ref{32}) becomes 
\begin{equation}
(E+\alpha s)^{2}=L^{2}c^{2}\big(\frac{ds}{d\theta }\big)%
^{2}+s^{2}L^{2}c^{2}+m^{2}c^{4}.  \label{34}
\end{equation}%
Next we differentiate Eq.~(\ref{34}) with respect to $\theta $ and divide
through by $ds/d\theta $ to obtain 
\begin{equation}
\frac{d^{2}s}{d\theta ^{2}}+\Big[1-\big(\frac{\alpha }{Lc}\big)^{2}\Big]s-%
\frac{E\alpha }{L^{2}c^{2}}=0.  \label{35}
\end{equation}%
Equation~(\ref{35}) is a second-order linear differential equation and can
be easily solved. Then we substitute the solution into the first-order
differential equation (\ref{34}) to determine some of the integration
constants.

The connection between position and time can be obtained from Eq.~(\ref{17})
for $\dot{r}=dr/dt$, 
\begin{equation}  \label{36}
t={\displaystyle\int\limits_{r_{0}}^{r}}\frac{dr}{c^{2}}\big( 1-\big[ 1+%
\frac{L^{2}}{m^{2}r^{2}c^{2}}\big]\big[ \frac{mc^{2}}{E+(\alpha/r)}\big]^{2}%
\big)^{-1/2}.
\end{equation}
Alternatively, the angular momentum equation can be combined with the
orbital equation. Thus Eqs.~(\ref{eq:relL}) and (\ref{eq:ee}) can be
combined to give 
\begin{equation}
\frac{r^{2}\dot{\theta}}{c^{2}}=\frac{L}{E+\alpha/r},
\end{equation}
so that from $\dot{\theta}=d\theta/dt$, we have 
\begin{equation}  \label{38}
t={\displaystyle\int\limits_{\theta_{0}}^{\theta}}\frac{r^{2}d\theta(E+%
\alpha/r)}{Lc^{2}},
\end{equation}
where $r(\theta)$ is to be evaluated from the orbital equations.

\section{Solution of the Orbital Equations}

We obtain different solutions of Eqs.~(\ref{34}) and (\ref{35}) depending on
the value of the angular momentum $L$. If $L>\alpha/c$, we have 
\begin{equation}  \label{39}
s=\frac{1}{r}=\sqrt{\frac{E^{2}L^{2}c^{2}-m^{2}c^{4}(L^{2}c^{2}-\alpha^{2})}{%
(L^{2}c^{2}-\alpha^{2})^{2}}}\cos\big[ \sqrt{1-\big( \frac{\alpha}{Lc}\big)%
^{2}}(\theta-\theta_{0})\big] +\frac{E\alpha}{L^{2}c^{2}-\alpha^{2}}.
\end{equation}
If $L=\alpha/c$, we have 
\begin{equation}
s=\frac{1}{r}=\frac{1}{2}\big( \frac{E}{\alpha}\big) (\theta-\theta_{0})^{2}+%
\frac{m^{2}c^{4}-E^{2}}{2E\alpha}.
\end{equation}
For $L<\alpha/c$, we have 
\begin{equation}
s=\frac{1}{r}=\sqrt{\frac{m^{2}c^{4}(\alpha^{2}-L^{2}c^{2})+E^{2}L^{2}c^{2}}{%
(\alpha^{2}-L^{2}c^{2})^{2}}}\cosh\big[ \sqrt{\big( \frac{\alpha}{Lc}\big)%
^{2}-1}(\theta-\theta_{0})\big]-\frac{E\alpha}{\alpha^{2}-L^{2}c^{2}}.
\end{equation}
Finally, if $L=0$, the orbit is a straight line $\theta=\theta_{0}$.

In the nonrelativistic limit $1/c\rightarrow 0$, only the first of these
solutions ($L>\alpha /c$) does not degenerate into a straight-line orbit.
The relativistic orbits in Eq.~(\ref{39}) are the bound rosettes and
unbounded scattering orbits which usually appear in textbooks.\cite%
{Goldstein} Thus if we set $E=\mathcal{E}+mc^{2}$ and take the limit $%
1/c\rightarrow 0$ in Eq.~(\ref{39}), we find 
\begin{equation}
s=\frac{1}{r}=\frac{m\alpha }{L_{nr}^{2}}\sqrt{1+\frac{2L_{nr}^{2}\mathcal{E}%
_{nr}}{m\alpha ^{2}}}\cos (\theta -\theta _{0})+\frac{m\alpha }{L_{nr}^{2}},
\end{equation}%
which is the standard nonrelativistic result for the hyperbolas, parabolas,
and ellipses of Kepler orbits.

The various types of orbits are sketched in Figs.~1--5, showing bound and
unbound relativistic trajectories. The unbound orbits, shown in Figs.~1--3,
have a total relativistic energy at least as large as the particle rest
energy, $E=\mathcal{E}+mc^{2}\geq mc^{2}$. These orbits extend to spatial
infinity in at least one of their time limits. For $L>\alpha /c$, these
orbits are familiar scattering orbits as shown in Fig.~1, where the orbit
looks like a parabolic or hyperbolic nonrelativistic orbit. However, for $%
L>\alpha /c$ but close to $\alpha /c$, we are reminded that relativity
changes the familiar nonrelativistic scattering orbits. In Fig.~2 we see
that the scattering trajectory loops around the center of the potential
before receding to infinity. These loops do not occur for the parabolic or
hyperbolic orbits of nonrelativistic scattering. As $L$ decreases and
becomes closer to $\alpha /c$, the number of times the orbit loops around
the potential center increases. When $L\leq \alpha /c$, the character of the
trajectory changes. The orbital looping now continues all the way to the
center of the potential; the unbound orbit extends to infinity in only one
time direction and spirals into the potential center in the other (see
Fig.~3). In contrast, the parabolic and hyperbolic orbits of nonrelativistic
mechanics never reach the center of the potential, unless $L=0$.

The bound orbits have total energy $E$ smaller than the particle rest
energy, $E=\mathcal{E}+mc^{2}<mc^{2}$, corresponding to negative values of $%
\mathcal{E}=E-mc^{2}$. In nonrelativistic mechanics, all the bound orbits
are ellipses unless $L_{nr}=0$. For $L>\alpha /c$, the relativistic orbits
take the rosette shape of precessing ellipses shown in Fig.~4. These orbits
have two turning points. \ The orbits are familiar as the Sommerfeld
relativistic orbits of old quantum theory,\cite{Ruark} and reduce to the
familiar ellipses in the nonrelativistic limit. For $L\leq \alpha /c$, the
relativistic orbits spiral out from the center of the $1/r$ potential and
back into the center at early and late times as shown in Fig.~5. In the
nonrelativistic limit, these orbits become straight-line trajectories for $%
L_{nr}=0$.

From Eqs.~(\ref{36}) and (\ref{38}), we find that all the trajectories
involve finite time changes for finite changes in $r$ or $\theta $. For the
trajectories with $L\leq \alpha /c$ (which spiral into the potential
center), the particles arrive at the potential center in a finite time
starting from a finite radius. This result can be seen from Eq.~(\ref{36})
which connects the change in the radius $r$ with the elapsed time. For $%
L=\alpha /c$, an expansion for small $r$ gives an integral of the form 
\begin{equation}
t=\!{\displaystyle\int\limits_{0}}dr\Big(\frac{1}{2c^{2}}\sqrt{\frac{2\alpha 
}{Er}}+\mathcal{O}(\sqrt{r})\Big),
\end{equation}%
which is well behaved at the lower limit. If $L<\alpha /c$, an expansion for
small $r$ gives 
\begin{equation}
t=\!{\displaystyle\int\limits_{0}}\!dr\Big(\frac{1}{c^{2}\sqrt{%
1-L^{2}c^{2}/\alpha ^{2}}}+\mathcal{O}(r)\Big).
\end{equation}%
The square root is real because of the condition $L<\alpha /c$, and again
the integral is finite. Thus all the trajectories involve only finite times
in the vicinity of the $1/r$ potential center.

\section{Relevance of the Trajectories to Current Research}

The existence of classical relativistic scattering trajectories that spiral
into the center of the $1/r$ potential while preserving energy and angular
momentum was pointed out by Darwin\cite{Darwin} in 1913. Darwin's analysis
was made in connection with electron scattering based on Rutherford's model
of the nuclear atom. In particular, Darwin was concerned with the spiral
trajectories and suggested that they should not occur in nature because they
would cause chemical transmutation. He speculated that ``There must
therefore be some way by which the electron can escape from the extreme
neighborhood of the nucleus."\cite{Darwin} He suggested that the presence of
these unobserved spiral trajectories indicated a further failure of
classical electromagnetic theory to describe experimental observation.
Darwin's work was published just before Bohr proposed his stationary-state
model of the atom.

Although Darwin was concerned primarily with the Rutherford scattering
problem of relativistic mechanics, he estimated the radiative corrections
for his scattering trajectories. It is interesting that a relativistic
treatment of the old classical problem of radiative atomic collapse gives a
different result from the familiar nonrelativistic treatment. In both the
nonrelativistic and relativistic analyses, the radiative loss of mechanical
energy as a charged classical particle moves in a Coulomb potential is such
as to make the orbit more circular. In the nonrelativistic treatment, the
charged particle radiates an infinite amount of energy as it spirals into
the center at ever higher speed and ever lower energy $\mathcal{E}%
_{nr}=-Ze^{2}/2r\rightarrow -\infty $. However, in the relativistic case,
the radiative loss of energy ends when the particle speed in the circular
orbit has reached the limit $v\rightarrow c$, the angular momentum in Eq.~(%
\ref{13}) has reached $L\rightarrow Ze^{2}/c$, the radius in Eq.~(\ref%
{eqrelv}) has shrunk to $r\rightarrow 0$, and the total energy in Eq.~(\ref%
{25}) has reached $E\rightarrow 0$. Thus a charged particle of rest energy $%
mc^{2}$ that starts at a large radius and small velocity with total energy $%
E=mc^{2}$ will radiate away the energy $mc^{2}$, and not the divergent
radiation energy found in the nonrelativistic treatment.

Today, physicists are not concerned about Darwin's spiral trajectories nor
the finite radiation energy loss given by a relativistic treatment of atomic
collapse. Of course, most physicists are not aware of their existence,
because quantum mechanics has changed our views of the atom and has made
such awareness unnecessary. However, occasionally some researchers are
interested to see just how far one can push a classical or semi-classical
interpretation of atomic physics. It has been suggested\cite{delaP} that the
inclusion of classical electromagnetic zero-point radiation might allow an
understanding of electron motion as a sort of Brownian motion which avoids
the problem of atomic collapse. Thus an electron might indeed lose energy
through radiation while accelerating in the electric field of the nucleus,
but the electron might also pick up energy from the zero-point radiation.
The balance between these two effects might account for atomic structure.

Recently Cole and Zou\cite{Cole} have carried out simulations of the
classical hydrogen ground state assuming classical zero-point radiation.
They conclude that the electron neither falls into the nucleus nor is
ionized, but rather assumes a stationary probability distribution. This
ground state probability distribution approximates that given by the Schr%
\"{o}dinger equation. This agreement is remarkable and requires confirmation
and further understanding. However, in contrast to these computer simulation
results are various attempts to calculate the electron Brownian motion
within the basic zero-point model using nonrelativistic mechanics.\cite%
{analytic} These analytic calculations suggest that the electron would not
fall into the nucleus due to radiation emission, but rather would be ejected
from the atom (self-ionization) due to the absorption of excessive
zero-point energy in the plunging orbits of very small angular momentum.
However, as we have emphasized here, the nonrelativistic orbits of very
small angular momenta represent the region of failure of the nonrelativistic
approximation. For small angular momentum $L<\alpha /c$, the correct
relativistic orbits spiral in toward the nucleus while conserving the total
energy and angular momentum. This behavior will lead to a totally different
interaction with zero-point radiation from that assumed in the
nonrelativistic calculations.\cite{analytic} Hence, the validity of the
self-ionization conclusions in the analytic work are questionable.\cite%
{Boyer}

We see that nearly a century after Darwin's work, a knowledge of the orbits
of a relativistic particle in a $1/r$ potential again seems significant. An
understanding of the possibilities and limitations of a classical or
semi-classical treatment of hydrogen depends upon a correct treatment of the
mechanical trajectories of a relativistic particle in the Coulomb potential.

\section{Acknowledgment}

I wish to thank Professor Martin Ligare for his helpful comments on the
first version of this manuscript and for bringing Dr. Ulf Torkelsson's work
to my attention.

\section*{Figure Captions}

Fig.~1. Sketch of an unbound orbit of total energy $E=\mathcal{E}
+mc^{2}\geq mc^{2}$ and angular momentum $L >> \alpha/c$ in a $1/r$
potential centered on the origin. The orbit is similar to a parabolic or
hyperbolic nonrelativistic scattering orbit.

\bigskip Fig.~2. Sketch of an unbound orbit of energy $E\geq mc^{2}$ and
angular momentum $L$ slightly larger than $\alpha /c$ in a $1/r$ potential.
Because $L$ is only slightly larger than $\alpha /c$, the orbit makes loops
around the scattering center. Such loops do not occur for an unbounded
nonrelativistic orbit which is always a parabola or hyperbola.

\bigskip Fig.~3. Sketch of an unbound orbit of energy $E\geq mc^{2}$ and
angular momentum $L\leq \alpha /c$. One end of the orbit is at infinity and
the other corresponds to a spiral into or out of the center of the
potential. The nonrelativistic limit corresponds to a straight line orbit
with $L=0$.

\bigskip Fig.~4. Sketch of a bound orbit of energy $0<E<mc^{2}$ and angular
momentum $L> \alpha/c$. The orbit has two turning points and, in the
nonrelativistic limit, reduces to the familiar elliptical orbit of
nonrelativistic mechanics.

\bigskip Fig. 5. Sketch of a bound orbit of energy $E<mc^{2}$ and angular
momentum $L \leq \alpha/c$. The orbit ends in spirals to and from the
potential center. The nonrelativistic limit is a straight line orbit with $%
L=0$.


\begin{thebibliography}{9}
\bibitem{only} For no other potential of the form $V(r)=kr^{n}$, where $k$
and $n$ are constants, does relativity restrict the appearance of circular
orbits for non-zero angular momentum. \ This is easily found by direct
calculation analogous to that given in Eqs. (7)-(12).
\end{thebibliography}
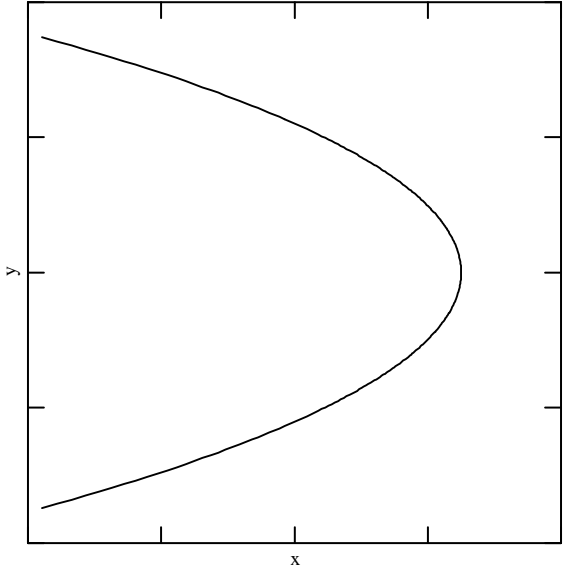
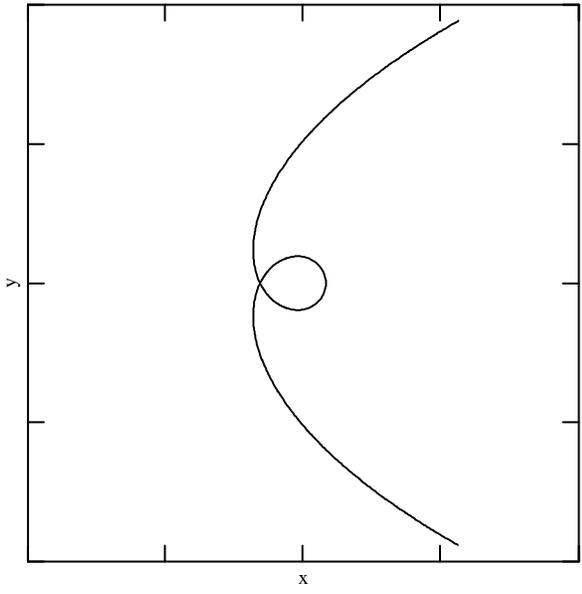
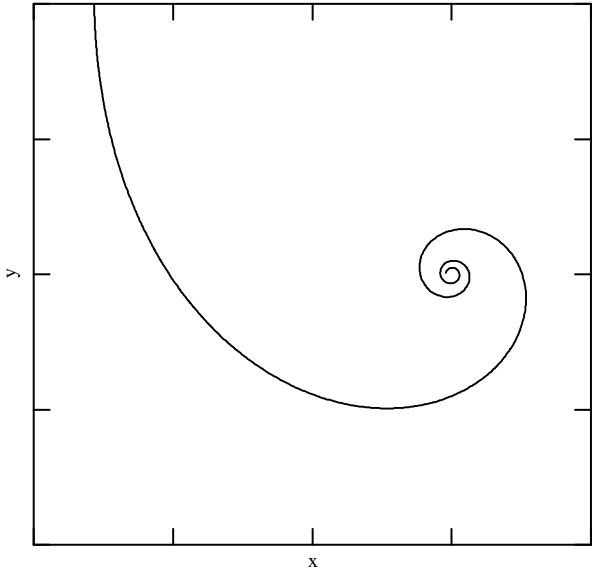
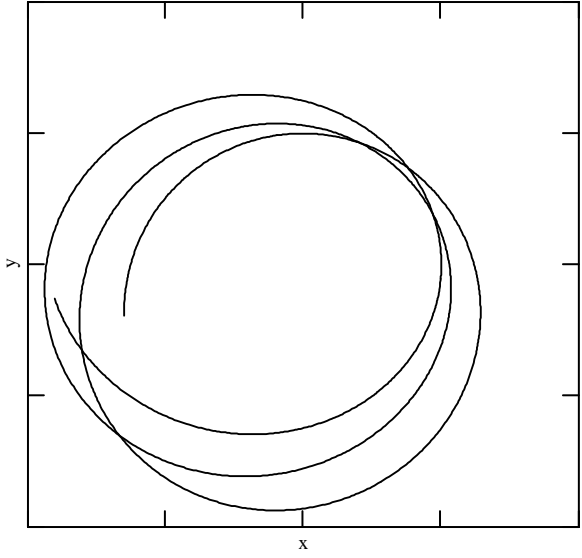
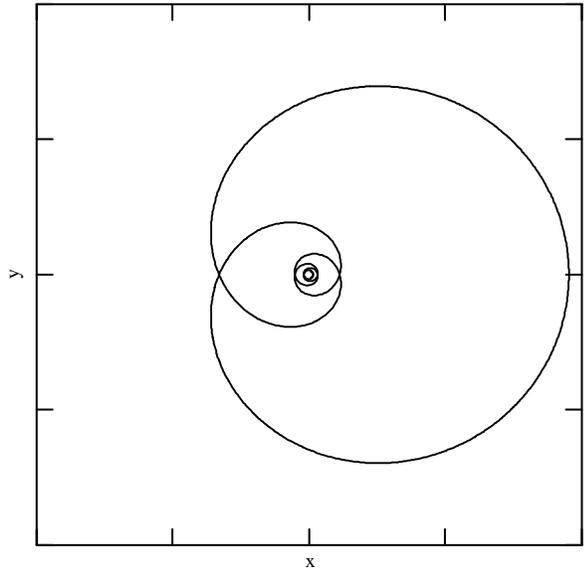
\end{document}